\documentclass[11pt]{article}
\usepackage{amssymb}
\usepackage{amsmath}
\usepackage[english]{babel}
\usepackage{graphicx}

\def\Journal#1#2#3#4{{#1} {\bf #2}, #3 (#4)}

\def\RNC{\em Rivista Nuovo Cimento}

\def\PLB{{\em Phys. Lett.}  B}
\def\PRL{\em Phys. Rev. Lett.}
\def\PRD{{\em Phys. Rev.} D}

\def\GaC{\em Gravitation and Cosmology}

\def\JETPL{\em JETP Lett.}

\def\CQG{\em Class. Quantum Grav.}

\def\MPLA{{\em Mod. Phys. Lett.}  A}
\def\IJTP{\em Int. J. Theor. Phys.}
\def\NJP{\em New J. of Phys.}
\def\JHEP{\em JHEP}
\def\EPHJ{\em Eur.Phys.J}

\def\s{{\,\rm s}}

\def\eV{\,{\rm eV}}
\def\keV{\,{\rm keV}}

\def\TeV{\,{\rm TeV}}

\def\({\left(}
\def\){\right)}
\def\cm{{\,\rm cm}}

\def\beq{\begin{equation}}
\def\eeq{\end{equation}}
\def\bea{\begin{eqnarray}}
\def\eea{\end{eqnarray}}

\begin{document}

    \begin{center}
        \large \textbf{Puzzles of Dark Matter - More Light on Dark Atoms?}
    \end{center}

    \begin{center}
   Maxim Yu. Khlopov$^{1,2,3}$, Andrey G. Mayorov $^{1}$, Evgeny Yu.
   Soldatov $^{1}$

    \emph{$^{1}$National Research Nuclear University "Moscow Engineering Physics Institute", 115409 Moscow, Russia \\
    $^{2}$ Centre for Cosmoparticle Physics "Cosmion" 115409 Moscow, Russia \\
$^{3}$ APC laboratory 10, rue Alice Domon et L\'eonie Duquet \\75205
Paris Cedex 13, France}

    \end{center}

\medskip

\begin{abstract}

Positive results of dark matter searches in experiments DAMA/NaI and
DAMA/LIBRA confronted with results of other groups can imply
nontrivial particle physics solutions for cosmological dark matter.
Stable particles with charge -2, bound with primordial helium in
O-helium "atoms" (OHe), represent a specific nuclear-interacting
form of dark matter. Slowed down in the terrestrial matter, OHe is
elusive for direct methods of underground Dark matter detection
using its nuclear recoil. However, low energy binding of OHe with
sodium nuclei can lead to annual variations of energy release from
OHe radiative capture in the interval of energy 2-4 keV in DAMA/NaI
and DAMA/LIBRA experiments. At nuclear parameters, reproducing DAMA
results, the energy release predicted for detectors with chemical
content other than NaI differ in the most cases from the one in DAMA
detector. Moreover there is no bound systems of OHe with light and
heavy nuclei, so that there is no radiative capture of OHe in
detectors with xenon or helium content.
Due to dipole Coulomb barrier, transitions to more energetic levels
of Na+OHe system with much higher energy release are suppressed in
the correspondence with the results of DAMA experiments. The
proposed explanation inevitably leads to prediction of abundance of
anomalous Na, corresponding to the signal, observed by DAMA.

\end{abstract}
\section{Introduction}
In our previous paper \cite{Levels} we have shown that the set of
conditions for dark matter candidates
\cite{book,Cosmoarcheology,Bled07}can be satisfied for new stable
charged particles, if they are hidden in neutral atom-like states.
To avoid anomalous isotopes overproduction, stable particles with
charge -1 (like tera-electrons \cite{Glashow,Fargion:2005xz}) should
be absent, so that stable negatively charged particles should have
charge -2 only. In the row of possible models, predicting such
particles
\cite{I,lom,Khlopov:2006dk,Q,Khlopov:2006dk,5,FKS,KK,Khlopov:2008rp}
stable charged clusters $\bar u_5 \bar u_5 \bar u_5$ of (anti)quarks
$\bar u_5$ of 5th family from the {\it spin-charge-family-theory}
\cite{Norma} can also find their place (see \cite{Discussion}).

In the asymmetric case, corresponding to excess of -2 charge
species, $X^{--}$, they bind in "dark atoms" with primordial $^4He$
as soon as it is formed in the Standard Big Bang Nucleosynthesis. We
call such dark atoms O-helium ($OHe$) \cite{I2} and assume that they
are the dominant form of the modern dark matter.

Here we concentrate on effects of O-helium dark matter in
underground detectors. We present qualitative confirmation of the
earlier guess \cite{I2,KK2,iwara,unesco} that the positive results
of dark matter searches in DAMA/NaI (see for review
\cite{Bernabei:2003za}) and DAMA/LIBRA \cite{Bernabei:2008yi}
experiments can be explained by O-helium, resolving the controversy
between these results and the results of other experimental groups.

\section{Radiative capture of OHe in the underground detectors}
\subsection{O-helium in the terrestrial matter} The evident
consequence of the O-helium dark matter is its inevitable presence
in the terrestrial matter, which appears opaque to O-helium and
stores all its in-falling flux.

After they fall down terrestrial surface, the in-falling $OHe$
particles are effectively slowed down due to elastic collisions with
matter. Then they drift, sinking down towards the center of the
Earth with velocity \beq V = \frac{g}{n \sigma v} \approx 80 S_3
A^{1/2} \cm/\s. \label{dif}\eeq Here $A \sim 30$ is the average
atomic weight in terrestrial surface matter, $n=2.4 \cdot 10^{24}/A \cm^{-3}$
is the number density of terrestrial atomic nuclei, $\sigma v$ is the rate
of nuclear collisions, $m_o \approx M_X+4m_p=S_3 \TeV$ is the mass of O-helium,
$M_X$ is the mass of the $X^{--}$ component of O-helium, $m_p$ is the mass of proton and $g=980~ \cm/\s^2$.

Near the Earth's surface, the O-helium abundance is determined by
the equilibrium between the in-falling and down-drifting fluxes.

The in-falling O-helium flux from dark matter halo is
$$
  F=\frac{n_{0}}{8\pi}\cdot |\overline{V_{h}}+\overline{V_{E}}|,
$$
where $V_{h}$-speed of Solar System (220 km/s), $V_{E}$-speed of
Earth (29.5 km/s) and $n_{0}=3 \cdot 10^{-4} S_3^{-1} \cm^{-3}$ is the
local density of O-helium dark matter. For qualitative estimation we
don't take into account here velocity dispersion and distribution of particles
in the incoming flux that can lead to significant effect.

At a depth $L$ below the Earth's surface, the drift timescale is
$t_{dr} \sim L/V$, where $V \sim 400 S_3 \cm/\s$ is given by
Eq.~(\ref{dif}). It means that the change of the incoming flux,
caused by the motion of the Earth along its orbit, should lead at
the depth $L \sim 10^5 \cm$ to the corresponding change in the
equilibrium underground concentration of $OHe$ on the timescale
$t_{dr} \approx 2.5 \cdot 10^2 S_3^{-1}\s$.

The equilibrium concentration, which is established in the matter of
underground detectors at this timescale, is given by
\begin{equation}
    n_{oE}=\frac{2\pi \cdot F}{V} = n_{0}\frac{n \sigma v}{4g} \cdot
    |\overline{V_{h}}+\overline{V_{E}}|,
\end{equation}
where, with account for $V_{h} > V_{E}$, relative velocity can be
expressed as
$$
    |\overline{V_{o}}|=\sqrt{(\overline{V_{h}}+\overline{V_{E}})^{2}}=\sqrt{V_{h}^2+V_{E}^2+V_{h}V_{E}sin(\theta)} \simeq
$$
$$
\simeq V_{h}\sqrt{1+\frac{V_{E}}{V_{h}}sin(\theta)}\sim
V_{h}(1+\frac{1}{2}\frac{V_{E}}{V_{h}}sin(\theta)).
$$
Here $\theta=\omega (t-t_0)$ with $\omega = 2\pi/T$, $T=1yr$ and
$t_0$ is the phase. Then the concentration takes the form
\begin{equation}
    n_{oE}=n_{oE}^{(1)}+n_{oE}^{(2)}\cdot sin(\omega (t-t_0))
    \label{noE}
\end{equation}

So, there are two parts of the signal: constant and annual
modulation, as it is expected in the strategy of dark matter search
in DAMA experiment \cite{Bernabei:2008yi}.

\subsection{Radiative capture of O-helium by sodium}

In the essence, our explanation of the results of experiments
DAMA/NaI and DAMA/LIBRA is based on the idea that OHe, slowed down
in the terrestrial matter and present in the matter of DAMA
detectors, can form a few keV bound state with sodium nuclei, in
which OHe is situated \textbf{beyond} the nucleus. Radiative capture
to this bound state leads to the corresponding energy release and
ionization signal, detected in DAMA experiments.

The rate of radiative capture of OHe by nuclei can be calculated
\cite{iwara,unesco} with the use of the analogy with the radiative
capture of neutron by proton with the account for: i) absence of M1
transition that follows from conservation of orbital momentum and
ii) suppression of E1 transition in the case of OHe. Since OHe is
isoscalar, isovector E1 transition can take place in OHe-nucleus
system only due to effect of isospin nonconservation, which can be
measured by the factor $f = (m_n-m_p)/m_N \approx 1.4 \cdot
10^{-3}$, corresponding to the difference of mass of neutron,$m_n$,
and proton,$m_p$, relative to the mass of nucleon, $m_N$. In the
result the rate of OHe radiative capture by nucleus with atomic
number $A$ and charge $Z$ to the energy level $E$ in the medium with
temperature $T$ is given by
\begin{equation}
    \sigma v=\frac{f \pi \alpha}{m_p^2} \frac{3}{\sqrt{2}} (\frac{Z}{A})^2 \frac{T}{\sqrt{Am_pE}}.
    \label{radcap}
\end{equation}

Formation of OHe-nucleus bound system leads to energy release of its
binding energy, detected as ionization signal.  In the context of
our approach the existence of annual modulations of this signal in
the range 2-6 keV and absence of such effect at energies above 6 keV
means that binding energy of Na-OHe system in DAMA experiment should
not exceed 6 keV, being in the range 2-4 keV. The amplitude of
annual modulation of ionization signal (measured in counts per day
per kg, cpd/kg) is given by
\begin{equation}
\zeta=\frac{3\pi \alpha \cdot n_o N_A V_E t Q}{640\sqrt{2}
A_{med}^{1/2} (A_I+A_{Na})} \frac{f}{S_3 m_p^2} (\frac{Z_i}{A_i})^2
\frac{T}{\sqrt{A_i m_p E_i}}= a_i\frac{f}{S_3^2} (\frac{Z_i}{A_i})^2
\frac{T}{\sqrt{A_i m_p E_i}}. \label{counts}
\end{equation}
Here $N_A$ is Avogadro number, $i$ denotes Na, for which numerical
factor $a_i=4.3\cdot10^{10}$, $Q=10^3$ (corresponding to 1kg of the
matter of detector), $t=86400 \s$, $E_i$ is the binding energy of
Na-OHe system and $n_{0}=3 \cdot 10^{-4} S_3^{-1} \cm^{-3}$ is the
local density of O-helium dark matter near the Earth. The value of
$\zeta$ should be compared with the integrated over energy bins
signals in DAMA/NaI and DAMA/LIBRA experiments and the result of
these experiments can be reproduced for $E_{Na} = 3 \keV$. The
account for energy resolution in DAMA experiments \cite{DAMAlibra}
can explain the observed energy distribution of the signal from
monochromatic photon (with $E_{Na} = 3 \keV$) emitted in OHe
radiative capture.

At the corresponding values of $\mu$ and $g^2$ there is no binding
of OHe with iodine and thallium \cite{Levels}.

It should be noted that the results of DAMA experiment exhibit also
absence of annual modulations at the energy of MeV-tens MeV. Energy
release in this range should take place, if OHe-nucleus system comes
to the deep level inside the nucleus. This transition implies
tunneling through dipole Coulomb barrier and is suppressed below the
experimental limits.

\subsection{OHe radiative capture by other nuclei}

For the chosen range of nuclear parameters, reproducing the results
of DAMA/NaI and DAMA/LIBRA, our results  \cite{Levels} indicate that
there are no levels in the OHe-nucleus systems for heavy nuclei. In
particular, there are no such levels in Xe and most probably in Ge,
what seem to prevent direct comparison with DAMA results in CDMS and
XENON100 experiments. However, even in this case presence of silicon
in the chemical composition of CDMS set-up can provide some
possibility for test of OHe interpretation of these results. The
levels in Si-OHe system were calculated in \cite{Levels}. The two
sets of solutions were obtained for each of approximation in
description of Yukawa potential:
\begin{itemize}
\item[i] the case (m) for nuclear Yukawa potential $U_{3m}$, averaged over the
orbit of He in OHe, 
\item[ii] the case (b) of the nuclear Yukawa potential
$U_{3b}$ with the position of He most close to the nucleus.
\end{itemize}
These two approximations correspond to the larger and smaller
distance effects of nuclear force, respectively, so that the true
picture should be between these two extremes.

For the parameters, reproducing results of DAMA experiment the
predicted energy level of OHe-silicon bound state is generally
beyond the range 2-6 keV, being in the most cases in the range of
30-40 keV or 90-110 keV by absolute value. It makes elusive a
possibility to test DAMA results by search for ionization signal in
the same range 2-6 keV in other set-ups with content that differs
from Na and I. Even in the extreme case (m) of ionization signal in
the range 2-6 keV our approach naturally predicts its suppression in
accordance with the results of CDMS \cite{Kamaev:2009gp}.

It should be noted that strong sensitivity of the existence of the
OHe-Ge bound state to the values of numerical factors \cite{Levels}
doesn't exclude such state for some window of nuclear physics
parameters. The corresponding binding energy  would be about 450-460
keV, what proves the above statement even in that case.

Since OHe capture rate is proportional to the temperature, it looks
like it is suppressed in cryogenic detectors by a factor of order
$10^{-4}$. However, for the size of cryogenic devices  less, than
few tens meters, OHe gas in them has the thermal velocity of the
surrounding matter and the suppression relative to room temperature
is only $\sim m_A/m_o$. Then the rate of OHe radiative capture in
cryogenic detectors is given by Eq.(\ref{radcap}), in which room
temperature $T$ is multiplied by factor $m_A/m_o$, and the
ionization signal (measured in counts per day per kg, cpd/kg) is
given by Eq.(\ref{counts}) with the same correction for $T$
supplemented by additional factors $2 V_h/V_E$ and
$(A_I+A_{Na})/A_i$, where $i$ denotes Si.
 To illustrate possible effects of OHe in
various cryogenic detectors we give in Tables~\ref{ta1} and
\ref{ta2} energy release, radiative capture rate and counts per day
per kg for the pure silicon for the preferred values of nuclear
parameters.

\begin{table}
\caption{Effects of OHe in pure silicon cryogenic detector in the
case m for nuclear Yukawa potential $U_{3m}$, averaged over the
orbit of He in OHe \cite{Levels}.}

\center
\begin{tabular}{|c|c|c|c|c|c|c|}
    \hline
        $g^2/\mu^2, GeV^{-1}$ & 242 & 242 & 257 & 257 & 395 & 395\\
    \hline
        Energy, $keV$ & 2.7 & 31.9 & 3.0 & 33.2 & 6.1 & 41.9\\
    \hline
        $\sigma V \cdot 10^{-33}, cm^3/s$ & 19.3 & 5.6 & 18.3 & 5.5 & 12.8 & 4.9\\
    \hline
        $\xi \cdot 10^{-2}, cpd/kg$ & 10.8 & 3.1 & 10.2 & 3.1 & 7.2 & 2.7\\
    \hline
\end{tabular}\label{ta1}

\end{table}

\begin{table}
\caption{Effects of OHe in pure silicon cryogenic detector for the
case of the nuclear Yukawa potential $U_{3b}$ with the position of
He most close to the nucleus \cite{Levels}.}

\center
\begin{tabular}{|c|c|c|c|c|c|c|}
    \hline
        $g^2/\mu^2, GeV^{-1}$ & 242 & 242 & 257 & 257 & 395 & 395\\
    \hline
        Energy, $keV$ & 29.8 & 89.7 & 31.2 & 92.0 & 42.0 & 110.0\\
    \hline
        $\sigma V \cdot 10^{-33}, cm^3/s$ & 5.8 & 3.3 & 5.7 & 3.3 & 4.9 & 3.0\\
    \hline
        $\xi \cdot 10^{-2}, cpd/kg$ & 3.3 & 1.9 & 3.2 & 1.9 & 2.7 & 1.7\\
    \hline
\end{tabular}\label{ta2}

\end{table}

\section{Conclusions}

The results of dark matter search in experiments DAMA/NaI and
DAMA/LIBRA can be explained in the framework of our scenario without
contradiction with the results of other groups. This scenario can be
realized in different frameworks, in particular, in the extensions
of Standard Model, based on the approach of almost commutative
geometry, in the model of stable quarks of 4th generation that can
be naturally embedded in the heterotic superstring phenomenology, in
the models of stable technileptons and/or techniquarks, following
from Minimal Walking Technicolor model or in the approach unifying
spin and charges. Our approach contains distinct features, by which
the present explanation can be distinguished from other recent
approaches to this problem \cite{Edward} (see also for review and
more references in \cite{Gelmini}).

The proposed explanation is based on the mechanism of low energy
binding of OHe with nuclei. Within the uncertainty of nuclear
physics parameters there exists a range at which OHe binding energy
with sodium is in the interval 2-4 keV. Radiative capture of OHe to
this bound state leads to the corresponding energy release observed
as an ionization signal in DAMA detector.

OHe concentration in the matter of underground detectors is
determined by the equilibrium between the incoming cosmic flux of
OHe and diffusion towards the center of Earth. It is rapidly
adjusted and follows the change in this flux with the relaxation
time of few minutes. Therefore the rate of radiative capture of OHe
should experience annual modulations reflected in annual modulations
of the ionization signal from these reactions.

An inevitable consequence of the proposed explanation is appearance
in the matter of DAMA/NaI or DAMA/LIBRA detector anomalous
superheavy isotopes of sodium, having the mass roughly by $m_o$
larger, than ordinary isotopes of these elements. If the atoms of
these anomalous isotopes are not completely ionized, their mobility
is determined by atomic cross sections and becomes about 9 orders of
magnitude smaller, than for O-helium. It provides their conservation
in the matter of detector. Therefore mass-spectroscopic analysis of
this matter can provide additional test for the O-helium nature of
DAMA signal. Methods of such analysis should take into account the
fragile nature of OHe-Na bound states, since their binding energy is
only few keV.

With the account for high sensitivity of the numerical results to
the values of nuclear parameters and for the approximations, made in
the calculations, the presented results can be considered only as an
illustration of the possibility to explain puzzles of dark matter
search in the framework of composite dark matter scenario. An
interesting feature of this explanation is a conclusion that the
ionization signal expected in detectors with the content, different
from NaI, should be dominantly in the energy range beyond 2-6 keV.

Moreover, it is shown that in detectors, containing light nuclei
(e.g. helium-3) and heavy nuclei (e.g. xenon) there should be no
bound states with OHe. In the framework of our approach it means
that the physical nature of effects, observed in DAMA/NaI and
DAMA/LIBRA experiments, cannot be probed in XENON10, XENON100
experiments or in the future detectors with He-3 content. Test of
the nature of these results in CDMS experiment should take into
account the difference in energy release and rate of radiative
capture of OHe by silicon as well as in Ge, if OHe-Ge bound state
does exist. The uncertainty in the existence of OHe-Ge bound state
makes problematic direct test of our model in pure germanium
detectors.  It should be noted that the excess of low energy events
reported in the CoGent experiment can be hardly explained by
radiative capture of OHe. Therefore test of results of DAMA/NaI and
DAMA/LIBRA experiments by other experimental groups can become a
very nontrivial task.

It is interesting to note that in the framework of our approach
positive result of experimental search for WIMPs by effect of their
nuclear recoil would be a signature for a multicomponent nature of
dark matter. Such OHe+WIMPs multicomponent dark matter scenarios
that naturally follow from AC model \cite{FKS} and from models of
Walking technicolor \cite{KK2} can be also realized as OHe
(dominant)+5th neutrino (sub-dominant) model in the framework of
{\it spin-charge-family-theory} \cite{Norma} (see
\cite{Discussion}).

The presented approach sheds new light on the physical nature of
dark matter. Specific properties of dark atoms and their
constituents are challenging for the experimental search. The
development of quantitative description of OHe interaction with
matter confronted with the experimental data will provide the
complete test of the composite dark matter model.


\end{document}